\documentclass[aps,twocolumn]{revtex4}

\usepackage{graphics}
\usepackage{epsfig}
\usepackage{amsmath}

\def\ov#1{\overline{#1}}

\def\vb#1{\mbox{\boldmath$#1$}}
\def\pd#1#2{\frac{\partial #1}{\partial #2}}

\def\wh#1{\widehat{#1}}
\def\bdot{\,\vb{\cdot}\,}
\def\btimes{\,\vb{\times}\,}

\def\bhat{\wh{{\sf b}}}

\def\cal#1{\mathcal{#1}}

\def\eq#1{\eqref{eq:#1}}

\newcommand{\bc}{\begin{center}}
\newcommand{\ec}{\end{center}}
\newcommand{\bt}{\begin{tabbing}}
\newcommand{\et}{\end{tabbing}} 
\newcommand{\be}{\begin{eqnarray*}}
\newcommand{\ee}{\end{eqnarray*}}
\newcommand{\bs}{\begin{slide}}
\newcommand{\es}{\end{slide}}

\begin{document}

\title{Momentum conservation in dissipationless reduced-fluid dynamics}

\author{A.~J.~Brizard}
\affiliation{Department of Chemistry and Physics \\ Saint Michael's College, Colchester, VT 05439, USA}

\begin{abstract}
The momentum conservation law for general dissipationless reduced-fluid (e.g., gyrofluid) models is derived by Noether method from a variational principle. The reduced-fluid momentum density and the reduced-fluid canonical momentum-stress tensor both exhibit polarization and magnetization effects as well as an internal torque associated with dynamical reduction. As an application, we derive an explicit gyrofluid toroidal angular-momentum conservation law for axisymmetric toroidal magnetized plasmas.
\end{abstract}

\begin{flushright}
September 16, 2010 \\
\end{flushright}

%\pacs{52.30.Gz, 52.65.Tt}

\maketitle

Nonlinear reduced-fluid models play an important role in our understanding of the complex dynamical behavior of strongly magnetized plasmas. These nonlinear reduced-fluid models, in which fast time scales (such as the compressional Alfv\'{e}n time scale) have been asymptotically removed, include the reduced magnetohydrodynamic equations \cite{RMHD_1,RMHD_2,RMHD_3}, the nonlinear gyrofluid equations \cite{Brizard_92,Brizard_Hahm}, and several truncated reduced-fluid models (such as the Hasegawa-Mima equation \cite{HM,Hazeltine_RMHD_HM} and the Hasegawa-Wakatani equations \cite{HW}). Because the space-time-scale orderings for these reduced-fluid models are compatible with the nonlinear gyrokinetic space-time-scale orderings 
\cite{Brizard_Hahm}, they provide a very useful complementary set of equations that yield simpler interpretations of low-frequency turbulent plasma dynamics in realistic geometries.

The self-regulation of anomalous transport processes by plasma flows in turbulent axisymmetric magnetized plasmas has been intensively investigated in the past decade. Because a strong coupling has been observed \cite{Wang_etal} between toroidal-momentum transport and energy transport in such plasmas, it is natural to investigate the link between these two global conservation laws through an application of the Noether method on a suitable Lagrangian density \cite{PLS}. The purpose of the present Letter is to focus its attention on a momentum conservation law derived from a general reduced-fluid model 
\cite{Brizard_GF} and then explicitly investigate the reduced toroidal angular-momentum transport in axisymmetric magnetic geometry derived from it.

The general variational formulation of nonlinear dissipationless reduced-fluid models is expressed in terms of a Lagrangian density ${\cal L}(\psi^{\alpha})$ as a function of the multi-component field
\begin{equation}
\psi^{\alpha} \;\equiv\; (\Phi, {\bf A}, {\bf E}, {\bf B}; n, {\bf u}, p_{\|}, p_{\bot}).
\label{eq:psi_def}
\end{equation} 
Here, the electromagnetic fields $({\bf E},{\bf B})$ are defined in terms of the electromagnetic potentials $(\Phi,{\bf A})$ as
\begin{equation}
{\bf E} \;\equiv\; -\,\nabla\Phi \;-\; c^{-1}\partial{\bf A}/\partial t \;\;\;{\rm and}\;\;\; {\bf B} \;\equiv\; \nabla\btimes{\bf A}
\label{eq:EB_phiA}
\end{equation} 
and the reduced-fluid moments $(n,{\bf u},p_{\|},p_{\bot})$ are used for each plasma-particle species (with mass $m$ and charge $q$). We note that the Lagrangian formalism does not accommodate higher-order fluid moments (e.g., heat fluxes) and, therefore, the issue of fluid closure is completely ignored \cite{Park_etal}. These higher-order moments, as well as dissipative effects, can be added after the dissipationless reduced-fluid equations are derived by variational method \cite{SSB} (although a variational procedure 
\cite{Peng,Brizard_2005} can be  used to include heat fluxes in the pressure evolution equations). 

We begin with the general reduced Lagrangian density
\begin{eqnarray}
{\cal L}(\psi^{\alpha}) & \equiv & {\cal L}_{\rm M}({\bf E}, {\bf B}) \;+\; {\cal L}_{\Psi}(\Phi, {\bf A}; n, {\bf u}) \nonumber \\
 &  &+\; {\cal L}_{\rm F}(n, {\bf u}, p_{\|}, p_{\bot}; {\bf E}, {\bf B}), 
\label{eq:L_def}
\end{eqnarray}
where the electromagnetic Lagrangian density is ${\cal L}_{\rm M} \equiv (|{\bf E}|^{2} - |{\bf B}|^{2})/8\pi$, the gauge-dependent interaction Lagrangian density (summed over particle species) is ${\cal L}_{\Psi} \equiv -\;\sum\,q\,n\;( \Phi - {\bf A}\bdot{\bf u}/c)$, and the reduced-fluid Lagrangian density ${\cal L}_{\rm F}$ depends on $({\bf E},{\bf B})$ only through the process of dynamical reduction \cite{Brizard_Vlasovia}.

The reduced plasma-electrodynamic equations associated with the reduced Lagrangian density \eq{L_def} are divided into either constraint equations or dynamical equations. The electromagnetic constraint equations are 
\begin{equation}
\nabla\bdot{\bf B} \;=\; 0 \;=\; \nabla\btimes{\bf E} \;+\; c^{-1}\partial{\bf B}/\partial t, 
\label{eq:EB_constr}
\end{equation}
which are satisfied by the representation \eq{EB_phiA}. The reduced-fluid constraint equations, on the other hand, are the continuity equation 
\begin{equation}
\pd{n}{t} \;=\; -\;\nabla\bdot\left(n\frac{}{}{\bf u} \right),
\label{eq:n_eq}
\end{equation}
and the Chew-Goldberger-Low (CGL) pressure equations \cite{CGL,RR}
\begin{eqnarray}
\pd{p_{\|}}{t} & = & -\;\nabla\bdot\left(p_{\|}\frac{}{}{\bf u} \right) \;-\; 2\,p_{\|}\;\bhat_{0}\bhat_{0}:\nabla{\bf u}, \label{eq:ppar_eq} \\
\pd{p_{\bot}}{t} & = & -\;\nabla\bdot\left(p_{\bot}\frac{}{}{\bf u} \right) \;-\; p_{\bot}\;\left({\bf I} - \bhat_{0}\bhat_{0}\right):\nabla{\bf u}, 
\label{eq:pper_eq}
\end{eqnarray}
associated with the CGL pressure tensor 
\begin{equation}
{\sf P} \;\equiv\; p_{\|}\;\bhat_{0}\,\bhat_{0} \;+\; p_{\bot}\;({\bf I} - \bhat_{0}\,\bhat_{0}), 
\label{eq:P_CGL}
\end{equation}
where ${\bf B}_{0} \equiv B_{0}\;\bhat_{0}$ denotes the quasi-static background magnetic field. The reduced-fluid velocity ${\bf u}$ appearing in 
Eqs.~\eq{n_eq}-\eq{pper_eq} will be determined from the variational principle
\begin{equation}
\int\;\delta{\cal L}\;d^{4}x \;=\; 0. 
\label{eq:RVP}
\end{equation}

The reduced Maxwell equations and the reduced-fluid momentum equation are the dynamical equations derived from the reduced variational principle
\eq{RVP}. First, the reduced Maxwell equations
\begin{eqnarray}
\nabla\bdot{\bf D} & = & 4\pi\;\varrho, \label{eq:D_eq} \\
\nabla\btimes{\bf H} \;-\; \frac{1}{c}\;\pd{{\bf D}}{t} & = & \frac{4\pi}{c}\;{\bf J}, \label{eq:H_eq}
\end{eqnarray}
are expressed in terms of the reduced charge density $\varrho \equiv -\,\partial{\cal L}_{\Psi}/\partial\Phi = \sum\,qn$ and the reduced current density ${\bf J} \equiv c\,\partial{\cal L}_{\Psi}/\partial{\bf A} = \sum\,qn\,{\bf u} $, while the reduced electromagnetic fields ${\bf D} \equiv 4\pi\,\partial{\cal L}/\partial{\bf E} = {\bf E} + 4\pi\,{\bf P}$ and ${\bf H} \equiv -\,4\pi\,\partial{\cal L}/\partial{\bf B} = {\bf B} - 4\pi\,
{\bf M}$ are expressed in terms of the reduced polarization and magnetization
\begin{equation}
\left( {\bf P},\; {\bf M} \right) \;\equiv\; \left( \pd{{\cal L}_{\rm F}}{{\bf E}},\; \pd{{\cal L}_{\rm F}}{{\bf B}} \right).
\label{eq:pol_mag}
\end{equation} 
Equations \eq{D_eq}-\eq{H_eq} can also be expressed as
\begin{eqnarray}
\nabla\bdot{\bf E} & = & 4\pi\;\left( \varrho \;-\frac{}{} \nabla\bdot{\bf P}\right), \label{eq:DE_eq} \\
\nabla\btimes{\bf B} - \frac{1}{c}\pd{{\bf E}}{t} & = & \frac{4\pi}{c}\left( {\bf J} + \pd{{\bf P}}{t} + c\;\nabla\btimes{\bf M}\right), \label{eq:HB_eq}
\end{eqnarray}
where $\varrho_{\rm pol} \equiv -\,\nabla\bdot{\bf P}$ denotes the polarization density, ${\bf J}_{\rm pol} \equiv \partial{\bf P}/\partial t$ denotes the polarization current, and ${\bf J}_{\rm mag} \equiv c\,\nabla\btimes{\bf M}$ denotes the magnetization current.

Next, the reduced-fluid momentum equation is
\begin{eqnarray}
n\;\frac{d{\bf p}}{dt} & = & qn\; \left( {\bf E} \;+\; \frac{{\bf u}}{c}\btimes{\bf B} \right) \;+\; n \left( \nabla K \;-\; \nabla{\bf u}\bdot{\bf p} \right) \nonumber \\
 &  &-\; \left( p_{\bot}\,\nabla\gamma_{\bot} + \frac{p_{\|}}{2}\,\nabla\gamma_{\|} \;+\; \nabla\bdot{\sf P}_{*} \right),
\label{eq:pmom_eq}
\end{eqnarray}
where $d/dt \equiv \partial/\partial t + {\bf u}\bdot\nabla$, the reduced-fluid kinetic energy $K$ and the reduced-fluid kinetic momentum ${\bf p}$ are
\begin{equation}
\left( \begin{array}{c}
K \\
{\bf p}
\end{array} \right) \;\equiv\; \left( \begin{array}{c} 
\partial{\cal L}_{\rm F}/\partial n \\
n^{-1}\;\partial{\cal L}_{\rm F}/\partial{\bf u}
\end{array} \right), 
\label{eq:L_nu}
\end{equation}
and the symmetric reduced pressure tensor 
\begin{equation}
{\sf P}_{*} \;\equiv\; p_{\|}\,\gamma_{\|}\,\bhat_{0}\bhat_{0} \;+\; p_{\bot}\,\gamma_{\bot}\,({\bf I} - \bhat_{0}\bhat_{0})
\label{eq:P_star}
\end{equation}
is defined in terms of the coefficients $\gamma_{\|} \equiv -\,2\,\partial{\cal L}_{\rm F}/\partial p_{\|}$ and $\gamma_{\bot} \equiv -\,\partial
{\cal L}_{\rm F}/\partial p_{\bot}$. We note that this pressure tensor generalizes the CGL pressure tensor \eqref{eq:P_CGL} and includes standard finite-Larmor-radius (FLR) corrections through $\gamma_{\bot} \neq 1$ \cite{SSB}. 

The reduced equations \eq{EB_constr}-\eq{pper_eq}, \eq{D_eq}-\eq{H_eq}, and \eq{pmom_eq} satisfy the reduced momentum conservation law \cite{Brizard_GF}
\begin{equation}
\pd{\vb{\Pi}}{t} \;+\; \nabla\bdot{\sf T} \;=\; \nabla^{\prime}\ov{{\cal L}}, 
\label{eq:momentum_t}
\end{equation}
where the reduced momentum density is
\begin{equation}
\vb{\Pi} \;\equiv\; \sum\; n\;{\bf p} \;+\; \frac{{\bf D}\btimes{\bf B}}{4\pi\,c},
\label{eq:pi_def}
\end{equation}
the reduced canonical momentum-stress tensor is
\begin{eqnarray}
{\sf T} & \equiv & \sum\;{\sf P}_{*} \;+\; \left( {\cal L}_{\rm F} \;-\; \sum\; \eta^{a}\;\pd{{\cal L}_{\rm F}}{\eta^{a}}\right) \;{\bf I} \nonumber \\
 &  &+\; \left[\; \frac{1}{8\pi} \left( |{\bf E}|^{2} \;+\; |{\bf B}|^{2} \right) \;-\; {\bf B}\bdot{\bf M} \right]\;{\bf I} \nonumber \\
 &  &-\; \frac{1}{4\pi}\; \left( {\bf E}\;{\bf E} \;+\frac{}{} {\bf B}\;{\bf B} \right) \nonumber \\
 &  &+\; \left[ \sum\;n\;{\bf u}\;{\bf p} \;-\; \left( {\bf P}\;{\bf E} \;-\frac{}{} {\bf B}\;{\bf M} \right) \right],
\label{eq:T_def}
\end{eqnarray}
and $\nabla^{\prime}\ov{{\cal L}}$ denotes the spatial gradient of the reduced Lagrangian density $\ov{{\cal L}} \equiv {\cal L} - {\cal L}_{\Psi}$ with the dynamical fields \eq{psi_def} held constant. We note that, while the first three terms in the canonical momentum-stress tensor \eq{T_def} are symmetric while the remaining terms (on the last line) are not. The antisymmetric part $({\sf T}_{\sf A})_{ij} \equiv \frac{1}{2}\,
(T_{ij} - T_{ji}) \equiv \frac{1}{2}\,\varepsilon_{ijk}\,\tau^{k}$ of the canonical momentum-stress tensor \eq{T_def} can be expressed in terms of the reduced {\it internal torque} density
\begin{equation} 
\vb{\tau} \;\equiv\; \sum\;n\,{\bf u}\btimes{\bf p} \;+\; \left( {\bf E}\btimes{\bf P} \;+\frac{}{} {\bf B}\btimes{\bf M} \right),
\label{eq:tau_def}
\end{equation}
which exhibits classical {\it zitterbewegung} effects \cite{Barut} associated with the decoupling of the reduced-fluid momentum ${\bf p} \neq m{\bf u}$ and the reduced-fluid velocity ${\bf u}$ and the reduced polarization and magnetization effects. Here, the {\it internal} degrees of freedom of a gyrofluid {\it particle} are associated with the fast gyromotion that has been eliminated by dynamical reduction.

The symmetry of the momentum-stress tensor is physically connected to the conservation of angular momentum \cite{PLS,PM_85}, i.e., conservation of the total angular momentum (including internal angular momentum) explicitly requires a symmetric momentum-stress tensor. Since the left side of 
Eq.~\eqref{eq:momentum_t} is invariant under the transformation \cite{McLennan} $\vb{\Pi}^{\prime} \equiv \vb{\Pi} + \nabla\bdot{\sf S}$ and 
${\sf T}^{\prime} \equiv {\sf T} - \partial{\sf S}/\partial t$, the second-rank antisymmetric tensor ${\sf S} \equiv \frac{1}{2}\,\varepsilon_{ijk}\,\sigma^{k}$ can be chosen so that ${\sf T}^{\prime} \equiv {\sf T}_{\sf S} \equiv \frac{1}{2} (T_{ij} + T_{ji})$ is symmetric, i.e., the antisymmetric part ${\sf T}_{\sf A} \equiv \partial{\sf S}/\partial t$ yields the reduced internal angular momentum equation
\begin{equation}
\pd{\vb{\sigma}}{t} \;\equiv\; \vb{\tau}, 
\label{eq:spin_tau}
\end{equation}
where $\vb{\sigma}$ denotes the {\it internal} (spin) angular momentum density, and the reduced momentum conservation law \eq{momentum_t} becomes
\begin{equation}
\pd{}{t} \left( \vb{\Pi} \;-\; \frac{1}{2}\;\nabla\btimes\vb{\sigma}\right) \;+\; \nabla\bdot{\sf T}_{\sf S} \;=\; 
\nabla^{\prime}\;\ov{\cal L},
\label{eq:momentum_spin_t}
\end{equation} 
where we used the identity $\nabla\bdot{\sf S} \equiv -\,\frac{1}{2}\,\nabla\btimes\vb{\sigma}$. 

By applying the Noether Theorem in axisymmetric tokamak geometry, where 
\begin{equation}
{\bf B}_{0} \;\equiv\; \nabla\varphi \btimes\nabla\psi \;+\; B_{0\varphi}(\psi)\,\nabla\varphi 
\label{eq:B0_tok}
\end{equation}
and the background scalar fields are independent of the toroidal angle $\varphi$ (i.e., $\partial^{\prime}\ov{{\cal L}}/\partial\varphi \equiv 0$), we obtain the reduced toroidal angular-momentum transport equation \cite{Brizard_GF}
\begin{equation}
\pd{}{t}\left(\Pi_{\varphi} \;-\frac{}{} \sigma_{z}\right) \;+\;\nabla\bdot\left({\sf T}\bdot\pd{{\bf x}}{\varphi} \right) \;=\; 0.
\label{eq:Pi_varphi}
\end{equation}
Here, the toroidal momentum density is
\begin{eqnarray}
\Pi_{\varphi} & = & \vb{\Pi}\bdot\pd{{\bf x}}{\varphi} \;\equiv\; \wh{\sf z}\bdot{\bf x}\btimes\vb{\Pi} \label{eq:Pi_phi_def} \\
 & = & \sum\;n\,p_{\varphi} \;+\; \frac{(1 + b_{\|})\,D^{\psi}}{4\pi\,c} \;+\; \frac{{\bf D}\btimes{\bf B}_{\bot}}{4\pi\;c}\bdot
\pd{{\bf x}}{\varphi},
\nonumber
\end{eqnarray}
where the perturbed magnetic field is ${\bf B} - {\bf B}_{0} \equiv b_{\|}\,{\bf B}_{0} + {\bf B}_{\bot}$ and we used the identity ${\bf B}_{0}\btimes\partial{\bf x}/\partial\varphi \equiv \nabla\psi$. 

A more useful expression for Eq.~\eq{Pi_varphi}, however, is obtained in terms of the magnetic-flux average $\langle\cdots\rangle \equiv 
{\cal V}^{-1}\oint\,{\cal J}\,(\cdots)\;d\theta\,d\varphi$ as
\begin{equation}
\pd{\langle\Pi_{\varphi}\rangle}{t} \;+\; \frac{1}{{\cal V}}\;\pd{}{\psi} \left( {\cal V}\frac{}{}\left\langle T^{\psi}_{\;\;\varphi}\right\rangle 
\right) \;=\; \langle\tau_{z}\rangle,
\label{eq:Pi_surface}
\end{equation}
where ${\cal J} \equiv (\nabla\psi\btimes\nabla\theta\bdot\nabla\varphi)^{-1} = ({\bf B}_{0}\bdot\nabla\theta)^{-1}$ is the Jacobian for the magnetic coordinates $(\psi,\theta,\varphi)$ and ${\cal V} \equiv \oint\,{\cal J}\;d\theta\,d\varphi$. In Eq.~\eq{Pi_surface}, $\langle\tau_{z}\rangle$ denotes the reduced internal torque density and the surface-averaged toroidal angular-momentum flux is
\begin{eqnarray}
T^{\psi}_{\;\;\varphi} & \equiv &\nabla\psi\bdot{\sf T}\bdot\pd{{\bf x}}{\varphi} \label{eq:T_psi_varphi} \\
 & = & \sum\,n u^{\psi}\,p_{\varphi} \;-\; \frac{1}{4\pi}\,(D^{\psi}\,E_{\varphi} + B_{\bot}^{\psi}H_{\varphi}),
\nonumber
\end{eqnarray}
where $B_{\bot}^{\psi} \equiv {\bf B}_{\bot}\bdot\nabla\psi$ denotes the $\psi$-component of the perpendicular component of the perturbed magnetic field (since ${\bf B}_{0}\bdot\nabla\psi \equiv 0$).

We now investigate the surface-averaged toroidal angular-momentum conservation law \eq{Pi_surface} by considering the gyrofluid Lagrangian density 
\cite{Brizard_2005,Brizard_NFLR}
\begin{eqnarray}
{\cal L} & = & \frac{1}{8\pi} \left( |{\bf E}_{\bot}|^{2} - |{\bf B}|^{2} \right) \nonumber \\
 &  &+\; \sum\left[\;\frac{1}{2}\,mn\; u_{\|}^{2} \;-\; \left( n\,{\cal K}_{\rho} \;+\frac{}{} {\cal P} \right) \;\right] \nonumber \\
 &  &+\; \sum\; qn \left[\; \frac{{\bf u}}{c}\bdot\left( {\bf A}_{0} + A_{\|\rho}\,\bhat_{0} \right) \;-\; \;\Phi_{\rho}\;\right],
\label{eq:Lag_red_rho}
\end{eqnarray}
where $u_{\|} \equiv {\bf u}\bdot\bhat_{0}$ denotes the gyrofluid velocity along the unperturbed (background) magnetic-field lines and the pressure tensor \eq{P_star} is ${\sf P}_{*} \equiv {\sf P}$ (i.e., $\gamma_{\|} = 1 = \gamma_{\bot}$) with ${\cal P} \equiv \frac{1}{2}\,{\rm Tr}({\sf P}) = 
p_{\bot} + p_{\|}/2$. In Eq.~\eqref{eq:Lag_red_rho}, the zero-Larmor-radius limit of the gyrocenter dynamical reduction \cite{Brizard_Hahm} introduces the nonlinear FLR-corrected potentials
\begin{equation}
\left( \begin{array}{c}
\Phi_{\rho} \\
\\
A_{\|\rho}
\end{array} \right) \;\equiv\; \left( \begin{array}{c}
\Phi \;-\; \vb{\rho}_{\bot}\bdot{\bf E}_{\bot} \\
\\
A_{\|} \;-\; \bhat_{0}\bdot\vb{\rho}_{\bot}\btimes{\bf B}_{\bot}
\end{array} \right),
\label{eq:PhiA_rho}
\end{equation}
which generate the nonlinear FLR-corrected electromagnetic fields
\begin{equation}
\left( \begin{array}{c}
{\bf E}_{\rho} \\
{\bf B}_{\bot\rho}
\end{array} \right) \;\equiv\; \left( \begin{array}{c}
-\,\nabla\Phi_{\rho} - c^{-1}\bhat_{0}\,\partial A_{\|\rho}/\partial t \\
\nabla\btimes( A_{\|\rho}\;\bhat_{0})
\end{array} \right),
\label{eq:EB_rho}
\end{equation}
and the low-frequency ponderomotive potential
\begin{equation}
{\cal K}_{\rho} \;\equiv\; \frac{1}{2}\;m\,\Omega_{0}^{2}\;|\vb{\rho}_{\bot}|^{2} \;\equiv\; \frac{m}{2}\;|{\bf U}_{\bot}|^{2},
\label{eq:Krho_def}
\end{equation}
which generates the generalized ponderomotive force density $\nabla\bdot{\sf P}_{\rho} \equiv \nabla\bdot{\sf P} + n\,\nabla {\cal K}_{\rho}$. These nonlinear FLR-corrected fields are expressed in terms of the gyrofluid displacement \cite{Brizard_NFLR}
\begin{equation}
\vb{\rho}_{\bot} \;=\; \frac{c}{B_{0}\Omega_{0}} \left( {\bf E}_{\bot} \;+\; \frac{u_{\|}}{c}\;\bhat_{0}\btimes{\bf B}_{\bot} \right) \;\equiv\;
\frac{\bhat_{0}}{\Omega_{0}}\btimes{\bf U}_{\bot}.
\label{eq:rho_def}
\end{equation}
The partial derivatives \eq{L_nu} of the gyrofluid Lagrangian density \eq{Lag_red_rho} yield the gyrofluid kinetic energy $K = m\;u_{\|}^{2}/2 + 
{\cal K}_{\rho}$, and the gyrofluid momentum
\begin{equation}
{\bf p} \;=\; m\; \left( u_{\|} \;+\; {\bf U}_{\bot}\bdot\frac{{\bf B}_{\bot}}{B_{0}} \right)\;\bhat_{0} \;\equiv\; m\,u_{\|}^{*}\;\bhat_{0}, 
\label{eq:p_red}
\end{equation}
where $u_{\|}^{*}$ defines the gyrofluid velocity along the perturbed magnetic-field lines. Lastly, the partial derivatives (\ref{eq:pol_mag}) yield the gyrofluid polarization and magnetization
\begin{eqnarray}
{\bf P} & \equiv & \sum\;qn\;\vb{\rho}_{\bot} \;=\; \sum\;mn\;\frac{c\bhat_{0}}{B_{0}}\btimes{\bf U}_{\bot}, \label{eq:pol_red} \\
{\bf M} & \equiv & \sum\;qn\;\vb{\rho}_{\bot}\btimes\frac{u_{\|}}{c}\,\bhat_{0} \;=\; \sum\;mn\;\frac{u_{\|}}{B_{0}}\;{\bf U}_{\bot}, \label{eq:mag_red}
\end{eqnarray}
which appear in the Maxwell equations \eq{D_eq}-\eq{H_eq}. Note that the reduced internal torque \eq{tau_def} for this gyrofluid model is expressed as
\begin{eqnarray}
\vb{\tau} & = & \sum\;\frac{mnc}{B_{0}}\;(u_{\|} - u_{\|}^{*}) \left( {\bf E}_{\bot} \;+\; \frac{u_{\|}}{c}\,\bhat_{0}\btimes{\bf B}_{\bot}\right) 
\nonumber \\
 &  &+\; \sum\;mn\,u_{\|}^{*}\;\left({\bf u}_{\bot} \;-\frac{}{} {\bf U}_{\bot}\right)\btimes\bhat_{0},
\label{eq:tau_gyro}
\end{eqnarray}
where ${\bf E}_{\bot}\btimes{\bf P} + {\bf B}_{\bot}\btimes{\bf M} \equiv 0$.

The gyrofluid momentum equation \eq{pmom_eq} derived from the gyrofluid Lagrangian density \eq{Lag_red_rho} is expressed as
\begin{equation}
mn\,\bhat_{0}\;\frac{du_{\|}}{dt} \;=\; qn\;\left( {\bf E}_{\rho} \;+\; \frac{{\bf u}}{c}\btimes{\bf B}_{\rho}^{*} \right) \;-\; 
\nabla\bdot{\sf P}_{\rho},
\label{eq:EP_red}
\end{equation}
where ${\bf B}_{\rho}^{*} \equiv {\bf B}_{0} + u_{\|}\,(B_{0}/\Omega_{0})\;\nabla\btimes\bhat_{0} + {\bf B}_{\bot\rho}$. Equation 
\eq{EP_red} contains both the gyrofluid parallel-force equation 
\begin{equation}
mn\;\frac{du_{\|}}{dt} \;=\; {\sf b}_{\rho}^{*}\bdot\left( qn\;{\bf E}_{\rho} \;-\frac{}{} \nabla\bdot{\sf P}_{\rho} \right),
\label{eq:upar_dot}
\end{equation}
where ${\sf b}_{\rho}^{*} \equiv {\bf B}_{\rho}^{*}/B_{0}$, and the gyrofluid velocity
\begin{equation}
{\bf u} \;=\; u_{\|}\;{\sf b}_{\rho}^{*} \;+\; \left( qn\;{\bf E}_{\bot\rho} \;-\frac{}{} \nabla\bdot{\sf P}_{\rho} \right)\btimes
\frac{\bhat_{0}}{mn\,\Omega_{0}},
\label{eq:u_def}
\end{equation}
which includes the $E \times B$ velocity and the diamagnetic velocity and their nonlinear FLR corrections.

Next, we consider the gyrofluid version of the surface-averaged toroidal angular-momentum conservation law \eq{Pi_surface}, where the gyrofluid momentum 
\eq{p_red} is substituted in Eqs.~\eq{pi_def}-\eq{T_def}, with ${\sf P}_{*} = {\sf P}$ and $\eta^{a}\,\partial{\cal L}_{\rm F}/\partial\eta^{a} \equiv {\cal L}_{\rm F}$ in Eq.~\eq{T_def}. The gyrofluid version of the surface-averaged equation \eq{Pi_surface} is expressed in terms of the gyrofluid toroidal angular-momentum density \eq{Pi_phi_def}, with $p_{\varphi} = mu_{\|}^{*}\,b_{0\varphi}$ and
\begin{eqnarray}
D^{\psi} & \equiv & \left(1 + 4\pi\,\sum\frac{mnc^{2}}{B_{0}^{2}}\right)\,E^{\psi} \nonumber \\
 &  &+\; \left(4\pi\,\sum\frac{mn\,u_{\|}c}{B_{0}^{2}}\right)\;{\bf B}_{\bot}\bdot(\nabla\psi\btimes\bhat_{0}),
\label{eq:D_psi_def}
\end{eqnarray}
while the expression for Eq.~\eq{T_psi_varphi} is
\begin{eqnarray}
T^{\psi}_{\;\;\varphi} & = & \sum\,n \left( u^{\psi}\,p_{\varphi} \;-\; q\,\rho_{\bot}^{\psi}\,E_{\varphi} \;+\; mu_{\|}\;\frac{B_{\bot}^{\psi}}{B_{0}}
\;U_{\bot\varphi} \right) \nonumber \\
 &  &-\; \frac{1}{4\pi}\,\left( E^{\psi}\,E_{\varphi} \;+\frac{}{} B_{\bot}^{\psi}\,B_{\varphi}\right),
\label{eq:Tpsivarphi_gyro}
\end{eqnarray}
where we have separated the reduced polarization $P^{\psi} = \sum\,qn\,\rho_{\bot}^{\psi}$ and magnetization $M_{\varphi} = \sum\,mnu_{\|}
U_{\bot\varphi}/B_{0}$ from the Maxwell stress tensor.

As an application of the gyrofluid model \eqref{eq:Lag_red_rho}, we consider its electrostatic version (${\bf E} = -\,\nabla\Phi$, $u_{\|}^{*} = 
u_{\|}$, and ${\bf B} = {\bf B}_{0}$) and use the quasi-neutrality condition $\varrho \equiv \nabla\bdot{\bf P}$ [valid for $4\pi\,(\sum\,mnc^{2}/
B_{0}^{2}) \gg 1$]. The surface-averaged gyrofluid toroidal angular-momentum equation \eq{Pi_surface} therefore becomes
\begin{eqnarray}
 &  &\pd{}{t} \left( \left\langle \Pi_{\varphi\|}\right\rangle \;+\; \frac{1}{c}\,\langle P^{\psi}\rangle \;-\; \langle\sigma_{z}\rangle\right) 
\nonumber \\
 &  &=\; -\;\frac{1}{{\cal V}}\;\pd{}{\psi} \left[ {\cal V} \left( \left\langle \Gamma_{\varphi\|}^{\psi}\right\rangle \;+\; \left\langle
P^{\psi}\;\pd{\Phi}{\varphi} \right\rangle \right)\right],
\label{eq:Pi_varphi_Phi}
\end{eqnarray}
where $\Pi_{\varphi\|} \equiv (\sum\,mn\,u_{\|})\,b_{0\varphi}$, $\Gamma_{\varphi\|}^{\psi} \equiv (\sum\,mn\,u_{\|}u^{\psi})\,b_{0\varphi}$, and 
$P^{\psi} \equiv (\sum\,mnc^{2}/B_{0}^{2})\,E^{\psi}$. This equation was recently obtained \cite{Scott} (without the internal angular momentum 
$\sigma_{z}$) by direct evaluation of the time evolution of the surface-averaged gyrocenter moment $\langle\Pi_{\varphi}^{\rm can}\rangle \equiv \langle\Pi_{\varphi\|}\rangle - (\psi/c)\,\langle\varrho\rangle$ of the toroidal canonical momentum $mv_{\|}\,b_{0\varphi} - q\,\psi/c$, where the surface-averaged gyrofluid charge density $\langle\varrho\rangle = {\cal V}^{-1}\;\partial( {\cal V}\,\langle P^{\psi}\rangle)/\partial\psi$ is expressed in terms of the surface-averaged polarization component $\langle P^{\psi}\rangle$.

In this Letter, we have derived an exact toroidal angular-momentum conservation law \eq{Pi_varphi} that clearly exhibits the role played by the reduced internal torque $\tau_{z} \equiv \partial\sigma_{z}/\partial t$ in possibly driving spontaneous toroidal rotation in axisymmetric tokamak plasmas.

The Author greatly benefited from discussions with John A.~Krommes. This work was supported by a U.~S.~Department of Energy grant No.~DE-FG02-09ER55005.

\end{document}